\documentstyle[12pt]{article}
\input{epsf.tex}
\begin{document}
\vskip-2truecm{\hfill {\small DF/IST-3.2001  }} \\
\vskip-0.6cm\hfill {\small astro-ph/0101500  } \\
\begin{center}
{\huge\bf Linear perturbations in a
universe with a cosmological constant} \\
Ant\'onio  Vale\footnote{E-mail: l42550@alfa.ist.utl.pt} \\
{\scriptsize  Departamento de F\'{\i}sica,
	      Instituto Superior T\'ecnico,} \\

{\scriptsize  Av. Rovisco Pais 1, 1096 Lisboa, Portugal.} \\

Jos\'e P. S. Lemos\footnote{E-mail: lemos@kelvin.ist.utl.pt} \\
{\scriptsize  CENTRA, Departamento de F\'{\i}sica,
	      Instituto Superior T\'ecnico,} \\
{\scriptsize  Av. Rovisco Pais 1, 1096 Lisboa, Portugal, \&}
\\
{\scriptsize  Observat\'orio Nacional-MCT,} \\
{\scriptsize  Rua General Jos\'e Cristino 77, 20921 
Rio de Janeiro, Brazil.}
\end{center} 
\begin{abstract} 
There are now evidences that the cosmological constant $\Lambda$ has a
non-zero positive value.  Alternative scenarios to a pure
cosmological constant model are provided by quintessence, an effective
negative pressure fluid permeating the universe.  Recent results
indicate that the energy density $\rho$ and the pressure $p$ of this
fluid are constrained by $-\rho\leq p \buildrel<\over\sim-0.6\,\rho$.
Since $p=-\rho$ is equivalent to the pure cosmological constant model,
it is appropriate to analyze this particular, but important, case
further.
We study the linear theory of perturbations in a
Friedmann-Rober-tson-Walker universe with a cosmological constant.  We
obtain the equations for the evolution of the perturbations in the
fully relativistic case, for which we analyze the single-fluid and the
two-fluid cases. We obtain solutions to these equations in appropriate
limits. We also study the Newtonian approximation.  We find that for a
positive cosmological constant universe (i) the perturbations will
grow slower in the relativistic regime for a two-fluid composed of
dark matter and radiation, and (ii) in the Newtonian regime, the
perturbations stop growing.

\end{abstract}

\newpage

\section{Introduction}

Ever since Einstein first introduced it, the cosmological constant
$\Lambda$ has had a history full of ups and downs. The interest in
it has been reawakened by novel results, which now lead us to
believe that $\Lambda$ is an essential factor in the present
dynamical evolution of the universe. A result supporting the
existence of a cosmological constant is due to the recent
balloon-borne measurements of the cosmic background radiation
(\cite{boomerang,maxima}). These measurements lead to a non-zero
cosmological constant and to an $\Omega\simeq1$ universe.  Another
important result comes from the study of high redshift supernovae
(\cite{riess,perl}), which leads us to believe that the expansion of the
universe is accelerating. In combination, both studies yield for
the present energy densities of matter and the cosmological
constant roughly the values $\Omega_M\simeq0.3$,
$\Omega_{\Lambda}\simeq0.7$. An alternative to the cosmological 
constant scenario is provided by quintessence, a scalar field
furnishing an effective negative pressure fluid 
(\cite{peebles2,peeblesratra}). Recent results indicate that 
the energy density $\rho$ and the pressure $p$ of the fluid 
are constrained by $-\rho\leq p \buildrel<\over\sim -0.6\,\rho$ 
\cite{efstathiou}. 
Since $p=-\rho$ is 
equivalent to the pure cosmological constant model, it is appropriate 
to study this case further, in some detail, which we shall 
do in this paper. 

A long standing issue yet to be fully solved, in which the
cosmological constant $\Lambda$ (or variants such as 
quintessence) can play a role, is how the
multitude of structures we observe in the sky has formed from a
homogeneous background. In harmony with the novel findings
mentioned above, some works have also shown that the existence of
a positive cosmological constant in a cold dark matter scenario
yields results that fit better with observations, namely, the
power in the perturbation spectrum as a function of wavelength (or
mass) (\cite{kofmanetal,kraussturner,ostrikersteinhardt} , and the 
cluster abundance as a 
function of redshift \cite{bertschinger,bahcalletal}).
There are now some good reviews on the effect of a non-zero
cosmological constant on the universe
(\cite{carrolletal,cohn,sahnistarobinsky,carroll2000}).

One subject that has not been discussed in detail for a universe with
a positive cosmological constant, and which is a basic step in the
understanding of structure formation, is the study of the equations
and solutions of linear perturbations from the homogeneous
background. This has been done exhaustively for models considering
mainly dark matter, both warm and cold, normal matter and radiation
(see, e.g., \cite{peebles1,peebles3,padmanabhan1}, and references
therein), but mostly leaving out the existence of a possible
cosmological constant. Some exceptions are the papers by 
\cite{kofmanstaro} where
the authors study anisotropies in the microwavebackground in a
positive $\Lambda$ universe, \cite{silveira} where a perturbation
analysis and power spectrum calculation in a universe with a
time-dependent cosmological constant is performed, \cite{abramo} and
\cite{riazuelo} where one finds a perturbation analysis in models 
of quintessence, and \cite{proty} in which an extrapolation from a
$\Lambda=0$ case to a $\Lambda > 0$ case is made.

An excellent discussion on the theory of linear perturbations and
their connection to structure formation is given in the book of
Padmanabhan (\cite{padmanabhan1}). Our work presented here is
based mainly on chapter 4 of Padmanabhan's book. It is our aim to
extend the analysis and study of the linear perturbation theory by
including the effects of a positive cosmological constant. A
Newtonian analysis has been presented in (\cite{cohn}), which is
valid for perturbations with wavelengths much smaller than the
horizon size and for non-relativistic fluids. We enlarge the
analysis to the relativistic regime, valid for wavelengths larger
than the horizon size, and for relativistic fluids.  As it is
known, the existence of a positive $\Lambda$ affects the rate of
expansion of the universe, but not the the self-gravity of the
perturbations. This causes a suppression on the growth of the
perturbations. Since observations favour a flat universe we set
$\Omega=1$ in our study. We then find that, indeed, for a positive
$\Lambda$-Universe (i) the perturbations will grow slower in the
relativistic regime for a two-fluid composed of dark matter and
radiation (the growing mode grows with the power $1/4$ instead of
$1$ which characterizes the $\Lambda=0$ case), and (ii) in the
Newtonian regime, the perturbations stop growing in a
$\Lambda$-dominated universe (see also \cite{kofmanetal,cohn}).

Our paper is delineated as follows.  In section 2 the perturbed
equations for a single relativistic fluid are found and solved for
a radiation dominated phase, for a matter dominated phase and for
a $\Lambda$ dominated phase.  In section 3 the perturbed equations
for two relativistic fluids are studied and applied for the
coupled system of radiation and dark matter. In section 4 the perturbed
equations in the Newtonian approximation are obtained and solved.
These are valid for length scales smaller than the horizon size.
In section 5 we conclude.

\section{The relativistic case: single fluid}

\subsection{Equations}

When studying perturbations in general relativity there is always
the problem of fixing a gauge, or coordinate system. A discussion
of the perturbation
equations in different gauge conditions is given by Hwang and Noh
(\cite{Hwang}, see also \cite{bardeen}), while Unruh (\cite{Unruh})
gives a discussion on the problems of gauge fixing. We will fix
our gauge as the comoving
gauge, with the four-velocity $u^{\alpha}$ given by
$u^{\alpha}=(1,0,0,0)$, to derive the perturbation equations for a
nonzero $\Lambda$ universe. The derivation
of the equations for $\Lambda=0$ is given by Padmanabhan
(\cite{padmanabhan1}); we will here follow his procedure closely,
modifying the equations to introduce the cosmological constant.

We use the overdot to denote the derivative $u^{\alpha}D_{\alpha}$.
The projection tensor onto surfaces orthogonal to comoving world
lines, the comoving hypersurfaces, is
$h_{\alpha\beta}=g_{\alpha\beta}-u_{\alpha}u_{\beta}$. The natural
derivative on these comoving hypersurfaces is
$h_{\alpha}^{\beta}D_{\beta}$, and the Laplacian is given by
$\nabla^{2}=-h_{\alpha}^{\beta}D_{\beta}h^{\alpha\gamma}D_{\gamma}$.
The metric is the usual Friedmann-Robertson-Walker metric for a
flat universe, given by
\begin{equation} \label{FRW}
ds^2=dt^2-a(t)^2[dr^2+r^2(d\theta^2+\sin^2\theta d\phi^2)]\, .
\end{equation}
The energy momentum tensor is $T^{\alpha\beta}=
(\rho+p)u^{\alpha}u^{\beta}-p g^{\alpha\beta}$, where $\rho$ is
the energy density and $p$ the pressure of the fluid. We start with
the relativistic continuity and Euler equations, derived from
$D_{\alpha} T^{\alpha}_{\beta}=0$:
\begin{equation} \label{3.1}
\dot{\rho}=-3H(\rho+p) \, ,
\end{equation}
\begin{equation} \label{3.2}
\dot{u}_{\alpha}=\frac{h^{\beta}_{\alpha}D_{\beta}p}{\rho+p} \, .
\end{equation}
The Einstein field equations, $G_{\alpha\beta}-\Lambda
g_{\alpha\beta}=8\pi G T_{\alpha\beta}$, yield the Friedmann
equation
\begin{equation} \label{Fried}
H^{2}=\frac{8 \pi G}{3}\rho +\frac{\Lambda}{3} \, .
\end{equation}
Note that it is sometimes interesting to consider the cosmological
constant term, $\Lambda g^{\alpha\beta}$, as being an
energy-momentum tensor for a special fluid. Here we do not follow
this procedure, since we find more useful to see it as a
cosmological constant term.

Expanding $\dot{u}^{\alpha}$, using the definition for overdot
given above, we have
\begin{equation} \label{3.3}
D_{\alpha}\dot{u}^{\alpha}=(D_{\alpha}u^{\beta})(D_{\beta}u^{\alpha})
+u^{\beta}D_{\beta}(D_{\alpha}u^{\alpha})+u^{\beta}R_{\alpha\beta}u^{\alpha}
\, .
\end{equation}
In our chosen frame, the first two terms in the left hand side are
purely spatial. Using
\begin{equation} \label{3.4}
u^{\alpha}R_{\alpha\beta}u^{\beta}=R_{00}=4 \pi G(\rho+3p)-\Lambda
\, ,
\end{equation}
and, splitting the tensor $D_{i}u^{j}$
into a diagonal term (which in the absence of perturbations will
be related to the Hubble's constant $H$ by Hubble's Law) and
non-diagonal terms (which are perturbative terms whose square is
of second order) we get, up to first order
\begin{equation} \label{3.6}
D_{\alpha}\dot{u}^{\alpha}=3\dot{H}+3H^{2}+4\pi G(\rho+3p)-\Lambda
\, .
\end{equation}
Using now Euler's equation (\ref{3.2}), and again up to first
order, we can relate the left hand side of (\ref{3.6}) to
$\nabla^{2}p$ by
\begin{equation} \label{3.7}
D_{\alpha}\dot{u}^{\alpha}=-\frac{\nabla^{2}p}{\rho+p} \, .
\end{equation}
Combining equations (\ref{3.6}) and (\ref{3.7}), we have
\begin{equation} \label{3.8}
\dot{H}=-H^{2}-\frac{4\pi
G}{3}(\rho+3p)+\frac{\Lambda}{3}-\frac{\nabla^{2}p}{\rho+p} \, .
\end{equation}
Following Padmanabhan (\cite{padmanabhan1}), we need to change our
time variable to have a valid time label for the comoving
hypersurfaces. Taking $t$ to be the valid ordering label, we have:
\begin{equation} \label{3.9}
\frac{d\tau}{dt}=1-\frac{\delta p}{\rho+p}
\end{equation}
Taking now the perturbed quantities of the form $A=A_{b}+\delta
A$, where $A_b$ is the background, non-perturbed quantity and
$\delta A$ is a small perturbation, we can rewrite the continuity
equation (\ref{3.1}) up to first order as
\begin{equation} \label{3.10}
\frac{d\rho}{d\tau}=\dot{\rho}_{b}+\delta\dot{\rho}+\frac{\dot{\rho}_{b}}
{\rho_{b}+p_{b}}\,\delta p
\end{equation}
where the overdot is used to represent the derivatives with
respect to $t$. Taking the linearized right-hand side of
(\ref{3.1}), and equating the zeroth order terms, we get
\begin{equation} \label{3.11}
\dot{\rho}_{b}=-3H_{b}(\rho_{b}+p_{b}) \, ,
\end{equation}
and, for the first order terms,
\begin{equation} \label{3.12}
\delta\dot{\rho}=-3\delta H(\rho_{b}+p_{b})-3H_{b}\delta \rho \, .
\end{equation}
For $\dot{H}$, we have
\begin{equation} \label{3.13}
\frac{dH}{d\tau}=\dot{H}_{b}+\delta\dot{H}_{b}+
\frac{\dot{H}_{b}}{\rho_{b}+p_{b}}\delta
p \, .
\end{equation}
Linearizing equation (\ref{3.8}) and equating to equation
(\ref{3.13}), we have for the zeroth order terms,
\begin{equation} \label{3.14}
\dot{H}_{b}=-H_{b}^{2}-\frac{4\pi
G}{3}(\rho_{b}+3p_{b})+\frac{\Lambda}{3} \, .
\end{equation}
Using now the zeroth order expression (\ref{3.14}) and Friedmann's
equation, (\ref{Fried}), the first order term is
\begin{equation} \label{3.15}
\delta\dot{H}=-2H_{b}\delta H-\frac{4\pi
G}{3}\delta\rho-\frac{c_{\rm s}^{2}}{3}\frac{\nabla^{2}\delta\rho}{\rho_{b}+
p_{b}}
\, ,
\end{equation}
where we used the definition $\delta p=c_{\rm s}^{2}\delta\rho$, 
${c_{\rm s}}$ being the speed of sound. We can
then rewrite equation (\ref{3.12}) in first order in $\delta H$,
and differentiate it with respect to $t$ to get:
\begin{equation} \label{3.16}
\delta\dot{H}=-\frac{1}{3(1+w)}\Big[\ddot{\delta}-
3H_{b}(2w-c_{\rm s}^{2})\,\dot{\delta}+\frac{9}{2}H_{b}^{2}
(2c_{\rm s}^{2}-w+w^{2})\delta-\frac{3}{2}\Lambda w(1+w)\delta\Big] \, ,
\end{equation}
where $w=p/\rho$ and $\delta=\delta\rho/\rho$, and we have used
\begin{equation} \label{3.17}
\dot{w}=-3H_{b}(1+w)(c_{\rm s}^{2}-w) \, ,
\end{equation}
\begin{equation} \label{3.18}
\dot{H}_{b}=-\frac{3}{2}H_{b}^{2}(1+w)+\frac{1}{2}\Lambda(1+w) \,
.
\end{equation}
Using (\ref{3.12}), (\ref{3.15}) and (\ref{3.16}), we finally
arrive at: 
\begin{eqnarray} 
\label{3.19}
&\ddot{\delta}+H_{b}(2-6w+3c_{\rm s}^{2})\,\dot{\delta}-
\frac{3}{2}H_{b}^{2}(1-6c_{\rm s}^{2}-3w^{2}+8w)\delta &
\nonumber \\ &
+
\frac{1}{2}(1+w)(1-3w)\Lambda\delta=-
\Big(\frac{k\,c_{\rm s}}{a}\Big)^{2}\delta
\, ,&
\end{eqnarray}
where we introduced the Fourier transform such that
$\nabla^{2}\delta=-(k/a)^{2}\delta$. The Newtonian limit, is
obtained by taking $w\approx0$ and $c_{\rm s}^{2}\approx0$ in the left
hand side and using Friedmann's equation (\ref{Fried}) (see also
section \ref{secNewt}).

\subsection{Solutions}

\indent\indent We start by changing the independent variable in
equation (\ref{3.19}) from $t$ to $a$:
\begin{equation} \label{3.20}
a^2\frac{d^2\delta}{da^2}+A
a\frac{d\delta}{da}+\Big(B+\frac{k^2\,c_{\rm s}^{2}}{H^2a^2}\Big)
\delta=0 \, ,
\end{equation}
where
\begin{equation} \label{3.21}
A=\frac{3}{2}(1+2c_{\rm s}^{2}-5w)+\frac{3}{2}(1+w)\frac{\Lambda}{3H^2}\, ,
\end{equation}
\begin{equation} \label{3.22}
B=-\frac{3}{2}(1-6c_{\rm s}^{2}-3w^2+8w)+
\frac{3}{2}(1+w)(1-3w)\frac{\Lambda}{3H^2}
\, .
\end{equation}
\noindent \\ {\it Radiation dominated phase:}\\
\indent We take radiation to be the dominant component, and study its
perturbations $\delta_R$. We take $(k^2\,c_{\rm s}^{2})/(H^2a^2)\ll 1$, 
which in this
case corresponds to considering perturbations larger than the
Hubble radius. Taking $c_{\rm s}^{2}=w=1/3$, equation (\ref{3.20}) becomes
\begin{equation} \label{3.26}
a^2\frac{d^2\delta_R}{da^2}+2\frac{\Lambda}{3H^2}a\frac{d\delta_R}{da}
-2\delta_R=0 \, .
\end{equation}
We now rewrite Friedmann's equation as
\begin{equation} \label{friedmann}
H^2=\frac{8\pi G}{3}(\rho_R+\rho_{DM})+\frac{\Lambda}{3} \, .
\end{equation}
Ignoring the matter term in (\ref{friedmann}), we can write $H$ as
$H^2=H_0^2(\Omega_R\,a_0^4/a^4+\Omega_{\Lambda})$, where the quantities
$\Omega$ are the usual quotient of the density to the critical
density, given by $\Omega_R=\frac{8\pi G}{3 H_0^2}\rho_R$ and
$\Omega_{\Lambda}=\frac{\Lambda}{3H_0^2}$. Using this in
equation (\ref{3.26}), we get the solutions
\begin{equation} \label{3.27}
\delta_{R\,g}\propto
a^2\;_2F_1\Big[\frac{1}{4},1,\frac{7}{4},-\frac{\Omega_\Lambda}
{\Omega_R}\frac{a^4}{a_0^4}\Big] \, ,
\end{equation}
\begin{equation} \label{3.28}
\delta_{R\,d}\propto \frac{\sqrt{\Omega_\Lambda a^4+\Omega_R a_0^4}}{a}
\, ,
\end{equation}
where $d$ and $g$ stand for the decaying and growing modes and 
$_{2}F_{1}[a,b;c;z]$ is the hypergeometric function which can be 
given in terms of a 
power series $\sum_{k=0}^{\infty}(a)_k(b)_k/(c)_k \, z^k/k!$.
As $\Omega_\Lambda a^4 \ll \Omega_R a_0^4$, these solutions reduce
to those obtained in the $\Lambda=0$ case: $\delta_{R\,g}\propto
a^2\,,\,\delta_{R\,d}\propto 1/a$, as would be expect from the fact
that, in the radiation dominated phase, the cosmological constant
does not have much influence in the dynamics of expansion, given
by equation (\ref{Fried}).

To consider perturbations smaller than the Hubble radius in the
radiation dominated phase, we take 
$(k^2\,c_{\rm s}^{2})/(H^2a^2)\gg 1$, and
ignore the $B$ term in (\ref{3.20}). As we are considering the
radiation dominated phase, the cosmological constant does not have
much influence in the dynamics of evolution, and thus the solution
in this case is the same as the one for the $\Lambda=0$ case (see
\cite{padmanabhan1}).

\noindent \\ {\it Matter dominated phase:}\\
\indent We now study the perturbations in dark matter, $\delta_{DM}$
 in the matter
dominated phase (to study matter perturbations in the radiation
dominated phase, we need to consider the two fluid situation,
developed in the next section). We have $c_{\rm s}^2 =w=0$, and, using this
(but preserving $c_{\rm s}$ in the $(k^2\,c_{\rm s}^2)/(H^2a^2)$ term 
to account for pressure support), equation (\ref{3.20}) becomes
\begin{equation} \label{3.23}
a^2\frac{d^2\delta}{da^2}+\frac{3}{2}\Big(1+\frac{\Lambda}{3H^2}\Big)a
\frac{d\delta}{da}+\Big(\frac{k^2\,c_{\rm s}^2}{H^2a^2}-\frac{3}{2}+
\frac{3}{2}\frac{\Lambda}{3H^2}\Big)\delta=0 \,.
\end{equation}
We take $(k^2\,c_{\rm s}^2 )/(H^2a^2)\ll 1$, which here corresponds to
considering wavelengths larger than the Jeans wavelength,
$\lambda_{J}\, \propto \, c_{\rm s} / \rho^{1/2}$, which is the
characteristic scale at which pressure support becomes an
important factor.
For matter domination, we approximate (\ref{friedmann}) by taking
only the matter term. We can then ignore the $\Lambda$ terms in
(\ref{3.23}), and get the usual solutions (see, e.g.,
\cite{padmanabhan1}), given by $\delta_{DM}=Aa+Ba^{-3/2}$.

\noindent \\ {\it Lambda dominated phase:}\\
\indent If we take into account the cosmological constant, as well
as matter, we can write $H$ as $H^2=H_0^2
(\Omega_M\,a_0^3/a^3+\Omega_{\Lambda})$, where $\Omega_M=\frac{8\pi G}
{3H_0^2}\rho_{DM}$. Substituting this into
equation (\ref{3.23}) and once again ignoring the term
$(k^2\,c_{\rm s}^2 )/(H^2a^2)$, we will get the solutions:
\begin{equation} \label{2.23}
\delta_{DM\,d}=A \frac{\sqrt{1+\overline{x}^{3}}}{\overline{x}^{3/2}}
\end{equation}
\begin{equation}\label{2.24}
\delta_{DM\,g}=B\,\overline{x}\;\,_{2}F_{1}\Big[\frac{1}{3},1,\frac{11}{6},
-\overline{x}^{3}\Big]
\end{equation}
where $_2F_1$ is the hypergeometric function defined earlier, 
and $\overline{x}$ is
a modified scale factor defined by $\overline{x}=a/a_{\Lambda}=
x\,a_{{\rm eq}}/a_{\Lambda}$, where $x$ is the usual scale factor
$x=a/a_{{\rm eq}}$, with $a_{{\rm eq}}$ the scale factor at
the time of matter radiation
equivalence and $a_{\Lambda}=a_0(\Omega_M/\Omega_{\Lambda})^{1/3}$
the scale factor of the universe when the energy density of dark matter and
of the cosmological constant is the same. 
These solutions have also been obtained by Silveira and Waga \cite{silveira} 
for a study of perturbations in a universe with a time dependent 
cosmological constant. 
The behaviour for zero
cosmological constant is obtained by taking the limit $\overline{x}\ll 1$,
and corresponds to the solutions obtained for the matter dominated
phase. Solution (\ref{2.24}) is shown in Fig. 1. 
\vskip 1mm
\centerline{\epsffile{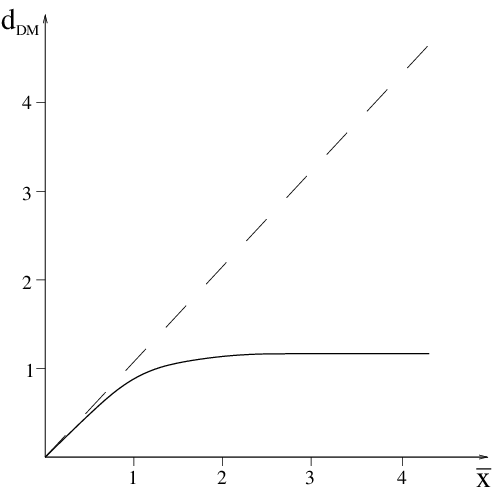}} 
{\noindent Figure 1. Amplitude of the dark matter perturbations as a function
of the modified scale factor $\overline{x}$; the dashed line
represents the evolution of the perturbation for $\Lambda=0$, the full
line in the presence of $\Lambda$. }
\vskip 2mm
One sees that for
$\overline{x} \ll 1$ the perturbations follow the behaviour for $\Lambda=0$,
 $\delta_{DM}\propto \overline{x}$, while for $\overline{x}\gg 1$, when the
cosmological constant dominates dynamically, the growth of perturbations
is suppressed.

We finally consider the case where the cosmological term dominates
in equation (\ref{Fried}). We can then write $H^2\simeq\Lambda/3$
and equation (\ref{3.23}) becomes simply
\begin{equation} \label{3.24}
a^2\frac{d^2\delta_{DM}}{da^2}+3\frac{d\delta_{DM}}{da}=0\, ,
\end{equation}
whose solution is
\begin{equation} \label{3.25}
\delta_{DM}=A+Ba^{-2} \, .
\end{equation}
Comparing this to the solution for the matter dominated phase,
given by $\delta=Aa+Ba^{-3/2}$, we can see that the presence of
the cosmological constant causes the perturbations to stop
growing.

The growing mode throughout the different phases is then given by:
\begin{displaymath} \label{3.29}
\left\{ \begin{array}{ll} \delta_R=a^2 & \textrm{(radiation
d.p., mode bigger than the H.r.)} \\
\delta_{DM}=a & \textrm{(matter d.p., mode bigger 
than J.l.)}
\\ \delta_{DM}=\textrm{constant} & \textrm{($\Lambda$ d.p.,
mode bigger than J.l.)}
\end{array} \right.
\end{displaymath}
where ``d.p.'' stands for dominated phase, ``H.r.'' for 
Hubble radius, and ``J.l.'' Jeans length. 
The one fluid relativistic description is good enough to study
perturbations in the dynamical dominant component. However, if one
wants to understand the behaviour of the other fluid components
one needs a two-fluid description. This we shall do now for the
case of radiation and dark matter.

\section{The relativistic case: two fluids}

\subsection{Equations}

\indent\indent In order to study simultaneously the perturbations
in radiation and dark matter, we shall first set up the equations
for the relativistic, multi-fluid case, and then specify for a
2-fluid with those components. We again follow
(\cite{padmanabhan1}), generalizing the procedure to include a
non-zero cosmological constant.\\ \indent We consider the case of
a perfect fluid made up of N perfect fluids, each with an energy
momentum tensor given by
$T_{N}^{\alpha\beta}=(\rho_{N}+p_{N})u_{N}^{\alpha}u_{N}^{\beta}-
p_{N}g^{\alpha\beta}$,
where, as before, we work in the comoving gauge,
${u}_{N}^\alpha\equiv(1,0,0,0)$. Each of these fluids will then
obey equations (\ref{3.1}) and (\ref{3.2}), where the quantities
will now have indexes $N$. Following the same procedure as in the
single fluid case, we get the following equations:
\begin{equation} \label{4.1}
\dot{\rho}_{N}=-3 H_{N} (\rho_{N}+p_{N}) \, ,
\end{equation}
\begin{equation} \label{4.2}
\frac{1}{3}D_{\alpha}\dot{u}_{N}^{\alpha}=\dot{H}_{N}+H_{N}^{2}+\frac{4\pi
G}{3}(\rho+p)-\frac{\Lambda}{3} \, ,
\end{equation}
\begin{equation} \label{4.3}
D_{\alpha}\dot{u}_{N}^{\alpha}=-\frac{\nabla^{2}p_{N}}{\rho_{N}+p_{N}}+
\frac{3(H-H_{N})}{\rho_{N}+p_{N}}\dot{p}_{N} \, .
\end{equation}
In equation (\ref{4.3}), the last term in the right hand side is new when
compared to equation (\ref{3.7}). This comes from the fact that, in the 
equation
for each fluid equivalent to (\ref{3.2}), we no longer have 
$h_{\alpha}^{\beta}$,
which is the correct projection tensor onto the comoving hypersurfaces, 
but a term
$h_{N\alpha}^{\phantom{N}\beta}=g_\alpha^\beta-u_{N\alpha}u_{N}^\beta$.
This new term reduces to zero in the
case of a single fluid. We also have for the total fluid
quantities:
\begin{equation} \label{4.4}
\rho=\sum_{N} \rho_{N} \, , \quad p=\sum_{N} p_{N} \, , \quad
H=\sum_{N} \frac{\rho_{N}+p_{N}}{\rho+p}H_{N} \, ,
\end{equation}
and
\begin{equation} \label{4.5}
H-H_{N}=\delta H - \delta H_{N} \, .
\end{equation}
\indent We can now follow the procedure of linearizing equations
(\ref{4.1}), (\ref{4.2}) and (\ref{4.3}), using the change in time
variable described in the last section, given by (\ref{3.9}), and
equation (\ref{4.5}), to get the following perturbation equations:
\begin{equation} \label{4.6}
\delta
\dot{\rho}_{N}=-3(\rho_{N}+p_{N})-3H\delta\rho_{N}-3H(\delta
p_{N}-\theta_{N} \delta p)
\end{equation}
\begin{equation} \label{4.7}
\delta \dot{H}_{N}=-2H_{N}\delta H_{N}-\frac{4\pi
G}{3}\delta\rho-\frac{1}{3}\frac{\nabla^{2}p_{N}}{\rho_{N}+p_{N}}+
\frac{\dot{p}_{N}}
{\rho_{N}+p_{N}}\bigg[\sum_{M}(\theta_{M}\delta H_{M})-\delta
H_{N}\bigg]
\end{equation}
where
\begin{equation} \label{4.8}
\theta_{N}=\frac{\rho_{N}+p_{N}}{\rho+p} \, .
\end{equation}
The quantities $w_{N}$ and ${c_{\rm s}}_{N}^{2}$ are as defined before for
the fluid $N$. For the total fluid these are given by:
\begin{equation} \label{4.9}
w=\frac{p}{\rho} \, ; \quad c_{\rm s}^{2} =\frac{\dot{p}}{\dot{\rho}} \, ,
\end{equation}
where $\rho$ and $p$ are defined by (\ref{4.4}). It is important
to notice that, unlike what happened with a single fluid, we have
here $c_{\rm s}^{2}\neq\delta p/\delta \rho$. The final equation for the
perturbations is then given by
{\setlength\arraycolsep{2pt}
\begin{eqnarray} \label{4.10}
\ddot{\delta}_N&+&(2+3{c_{\rm s}}_N^2-6w_N)H\dot{\delta}_N+3H\frac{1+w_N}{1+w}
{c_{\rm s}}_N^2\sum_M\frac{\rho_M}{\rho}
\Big(1-\frac{{c_{\rm s}}_M^2}{ {c_{\rm s}}_N^2}\Big)
\dot{\delta}_M\nonumber\\&+&
3\Big[\frac{1}{2}(-3H^2+\Lambda)({c_{\rm s}}_N^2-w_N)(1+w)
+5H^2({c_{\rm s}}_N^2-w_N)+2H{c_{\rm s}}_N
\dot{ {c_{\rm s}}        }_N\Big]\delta_N\nonumber\\&+&   
~9H^2\frac{1+w_N}{1+w}(c_{\rm s}^2-w){c_{\rm s}}_N^2\sum_M\frac{\rho_M}{\rho}
\Big(\frac{{c_{\rm s}}_M^2-w_M}{c_{\rm s}^2-w}-
\frac{{c_{\rm s}}_M^2}{ {c_{\rm s}}_N^2  }\Big)\delta_M
\nonumber\\ &-& \Big(\frac{6H^2}{1+w}+\frac{3}{2}(-3H^2+\Lambda)
\Big)(1+w_N)\sum_M\frac{\rho_M}{\rho} {c_{\rm s}}_N^2  
\delta_M \nonumber\\ &-&
3H\frac{1+w_N}{1+w}\sum_M
{c_{\rm s}}_M    
\frac{\rho_M}{\rho}\big(3H {c_{\rm s}}_M  (w-w_M)+
2\dot{  {c_{\rm s}}  }_M\big)-4\pi
G(1+w_N)\delta\nonumber\\&-&{c_{\rm s}}_N^2\nabla^2\delta_N=0
\end{eqnarray}}
\indent We are interested in studying the two fluid case, which in
this case will be radiation and dark matter. Following
(\cite{padmanabhan1}), we now introduce the adiabatic $\delta$ and
isocurvature $S$ perturbations related to the individual fluid
perturbations by:
\begin{equation} \label{4.11}
\delta=\frac{\rho_{1}
\delta_{1}+\rho_{2}\delta_{2}}{\rho_{1}+\rho_{2}}\, ; \quad
S=\frac{\delta_{1}}{1+w_{1}}-\frac{\delta_{2}}{1+w_{2}} \, .
\end{equation}
Using these definitions, the two perturbation equations become:
\begin{eqnarray} \label{4.12}
& \ddot{\delta}+H(2-6w+3c_{\rm s}^{2})\dot{\delta}-\frac{3}{2}H^{2}
(1-6c_{\rm s}^{2}+8w-3w^{2})\delta & \nonumber\\ & +\frac{1}{2}(1+w)(1-3w)
\Lambda\delta=-\frac{k^{2}}{a^{2}}\bigg(c_{\rm s}^{2}\delta+w
\eta\bigg) &
\end{eqnarray}
and
\begin{equation} \label{4.13}
\ddot{S}+H(2-3u^{2})\dot{S}=\frac{k^{2}}{a^{2}}\bigg(({c_{\rm s}}_1^2
- {c_{\rm s}}_2^2 )\frac{\delta}{1+w}-u^{2} S\bigg) \, ,
\end{equation}
where
\begin{equation} \label{4.14}
\eta=\frac{\rho_{1}(1+w_{1})\rho_{2}(1+w_{2})}{\rho
w(1+w)}( {c_{\rm s}}_1^2  - {c_{\rm s}}_2^2  )S \, ,
\end{equation}
\begin{equation} \label{4.15}
u^{2}=\frac{\rho_{1}(1+w_{1})     {c_{\rm s}}_2^2  +
\rho_{2}(1+w_{2}) {c_{\rm s}}_1^2 }{\rho(1+w)}
\, .
\end{equation}
Equation (\ref{4.12}) reduces to that of the single fluid case
(\ref{3.19}) if we take $\eta=0$ and a single fluid component.

\subsection{Solutions}

\indent\indent We discuss here the case of a two component system,
with radiation and dark matter. We will use as variable, instead
of $a$, $x=a/a_{{\rm eq}}$, with $a_{{\rm eq}}$ being the scale factor at the
time of matter-radiation equivalence. With this new variable we
can write:
\begin{equation} \label{4.16}
\frac{\rho_{DM}}{\rho_{{\rm eq}}}=\frac{1}{2x^{3}}\, ; \quad
\frac{\rho_{R}}{\rho_{{\rm eq}}}=\frac{1}{2x^{4}}\, ; \quad
\frac{\rho}{\rho_{{\rm eq}}}=\frac{\rho_{DM}+\rho_{R}}
{\rho_{{\rm eq}}}=\frac{1}{2x^{4}}(x+1)\, .
\end{equation}
Taking $w_{DM}=0={c_{\rm s}}_{DM}^2$ and $w_{R}=1/3={c_{\rm s}}_{R}^2 $, 
we have
\begin{equation} \label{4.17}
w=\frac{p_{DM}+p_{R}}{\rho_{DM}+\rho_{R}}=\frac{1}{3(1+x)} \, ;
\quad
c_{\rm s}^{2}=\frac{\dot{p_{DM}}+\dot{p_{R}}}{\dot{\rho_{DM}}+
\dot{\rho_{R}}}=
\frac{4}{9}\frac{1}{x+4/3}
\, .
\end{equation}
Using these definitions, we may write $H$ as
\begin{equation} \label{4.18}
H^{2}(x)=\frac{1}{2x^{4}}(x+1)H_{{\rm eq}}^{2}+\frac{\Lambda}{3}
\bigg(1-\frac{1}{2x^{4}}(x+1)\bigg)
\, .
\end{equation}
We will characterize the perturbations by the parameter $\omega$,
with
\begin{equation} \label{4.19}
\frac{k^{2}}{H^{2}a^{2}}=\frac{2x^{2}}{(1+x)}\omega^{2} \, ,
\end{equation}
with $\omega$ related to the wavelength by
$2\pi\omega=[d_{H}(t_{{\rm eq}})/\lambda(a_{{\rm eq}})]$, where $d_{H}$ is the
Hubble radius. Perturbations for which $\omega>1$ will enter the
Hubble radius in the radiation dominated phase, while for
$\omega<1$ they will enter it in the matter dominated phase.\\

Using the new variable $x$, and definitions (\ref{4.16})
to (\ref{4.19}), and defining $\hat{D}=x(d/dx)$, we can rewrite
equations (\ref{4.12}) and (\ref{4.13}) as
\begin{eqnarray} \label{4.20}
&\hat{D}^{2}\delta+\bigg[\frac{5}{2}\frac{x}{1+x}-
\frac{x}{x+4/3}-1+\frac{1}{2}\frac{\Lambda}
{3H^{2}}\bigg(3+\frac{1}{1+x}\bigg)\bigg]\hat{D}\delta
+\bigg(\frac{1}{2}\frac{x^{2}}{(1+x)^{2}}
+\frac{3x}{4} & \nonumber\\ &+
\frac{9}{4}\frac{x^{2}}{x+4/3}-\frac{3x^{2}}{1+x}-2+2x\frac{x+4/3}{(x+1)^{2}}
\frac{\Lambda}{3H^{2}}\bigg)\delta & \nonumber\\ &=
\frac{4}{3}\frac{\omega^{2}}{(x+1)(x+4/3)}\Big(x S-(x+1)\delta
\Big)\bigg[\frac{2}{3}\frac{x^{2}}{x+1}+\frac{1}{x^{2}}
\bigg(1-\frac{2x^{4}}{x+1}\bigg)
\frac{\Lambda}{3H^{2}}\bigg] &
\end{eqnarray}
\begin{eqnarray} \label{4.21}
& \hat{D}^{2}S+\bigg[\frac{x}{2(1+x)}-\frac{x}{x+4/3}+
\frac{1}{2}\bigg(3+\frac{1}{1+x}\bigg)
\frac{\Lambda}{3H^{2}}\bigg]\hat{D}S & \nonumber\\ &+
\frac{1}{3}\frac{\omega^{2}x}{x+4/3}\bigg[\frac{2x^{2}}{x+1}+
\bigg(\frac{1}{x^{2}}-\frac{2x^{2}}
{x+1}\bigg)\frac{\Lambda}{3H^{2}}\bigg]S & \nonumber\\ & =
\frac{1}{3}\frac{\omega^{2}}{x+4/3}\bigg[2x^{2}+\bigg(\frac{x+1}{x^{2}}
-2x^{2}\bigg)\frac{\Lambda}{3H^{2}}\bigg]\delta &
\end{eqnarray}
The perturbations $\delta$ and $S$ can be related to the
perturbations in the radiation and dark matter components by:
\begin{equation} \label{4.22}
\Delta_{R}=\frac{3}{4}\delta_{R}=\frac{x+1}{x+4/3}\delta-\frac{x}{x+4/3}S
\, ,
\end{equation}
\begin{equation} \label{4.23}
\Delta_{DM}=\delta_{DM}=\frac{x+1}{x+4/3}\delta+\frac{4}{3}\frac{1}{x+4/3}S
\, .
\end{equation}
With these relations we can now write the equations in
$\Delta_{R}$ and $\Delta_{DM}$ corresponding to equations
(\ref{4.20}) and (\ref{4.21}):
{\setlength\arraycolsep{2pt}
\begin{eqnarray} \label{4.24}
\bigg[\hat{D}^{2} &+& \bigg(\frac{1}{2}\,\frac{x}{x+1}+
\frac{3}{2}\,\frac{x+4/3}{x+1}
\frac{\Lambda}{3H^{2}}\bigg)\hat{D}+\frac{3}{2}\,
\frac{x}{x+1}\Big(\frac{\Lambda}
{3H^{2}}-1\Big)\bigg]\Delta_{DM} = \nonumber\\  &=&
\frac{4}{3}\,\frac{1}{x+4/3}\bigg[\hat{D}+2-\frac{x}{x+4/3}\bigg]\Delta_{R}
\nonumber\\&
-&
\frac{8}{9}\,\frac{1+x-2x^{4}}{x^{2}(x+1)(x+4/3)}\frac{\Lambda}{3H^{2}}
\omega^{2}\Delta_{R} \, , 
\end{eqnarray}
\begin{eqnarray} \label{4.25}
& \bigg[\hat{D}^{2}+\bigg(\frac{1}{2}\,\frac{x}{x+1}-1+\frac{3}{2}\,
\frac{x+4/3}{x+1}
\frac{\Lambda}{3H^{2}}\bigg)\hat{D}+\bigg(\frac{2}{3}\,
\frac{\omega^{2}x^{2}}{x+1}
+\frac{4}{3}\,\frac{1}{x+4/3}\Big(\frac{x}{x+4/3}-2\Big)\bigg)
& \nonumber\\ &+
\bigg(\frac{1}{x^{2}}-\frac{2}{3}\,\frac{1}{x(x+4/3)}-\frac{2x^{2}}{x+1}+
\frac{4}{3}\,
\frac{x^{3}}{(x+1)(x+4/3)}\bigg)\omega^{2}\frac{\Lambda}{3H^{2}}\bigg]
\Delta_{R} & \nonumber\\ &=
\Big[\frac{3}{2}\,\frac{x}{x+1}\bigg(1-\frac{\Lambda}{3H^{2}}\bigg)-
\frac{x}{x+4/3}\hat{D}\Big]\Delta_{DM} \, .&
\end{eqnarray}}
\indent We are interested in studying the solutions of these
equations in the two limiting cases $\omega^{2}\gg 1$ (perturbations
entering the Hubble radius in the radiation dominated phase) and
$\omega^{2}\ll 1$ (perturbations entergin the Hubble radius in the
matter dominated phase). We can further divide our study to the cases
where $x\ll 1$ and $x\gg 1$. As we have discussed earlier, the
effects of the presence of a cosmological constant will only be
important for $x\gg 1$, well into the matter dominated phase, as
the cosmological constant starts to assume a greater importance in
the dynamics of evolution. Therefore, we will here restrict
ourselves mainly to these cases; we will, however, discuss a
situation where $x\ll 1$ in order to certify that the solution
obtained is approximately the same as that of the case
$\Lambda=0$.

\noindent\\ \emph{$\omega^2 \gg 1$: perturbations entering the
Hubble radius in the radiation dominated phase}\\
The case $\omega^2 \gg 1$, for which the perturbations enter the
Hubble radius in the radiation dominated phase, has three distinct
cases: (i) when the mode is bigger than the Hubble radius, $\omega
x \ll 1$, the solution is very close to that of the $\Lambda=0$
case, given by $\Delta_{R}=\Delta_{DM}=x^2$, so we will not discuss it
further here; (ii) when $\omega x \gg 1$, but $x \ll 1$, the mode
has entered the Hubble radius, but the universe is still in the
radiation dominated phase; this case also shows very little
deviation from the zero cosmological constant case, but we will
consider it here to show this; (iii) finally, there is the case
$\omega x \gg 1$, and $x \gg 1$, when the universe is in the
matter dominated phase; this case shows significant deviation from
the $\Lambda=0$ case. For zero cosmological constant,
(\cite{naypad}) give an interesting approach to this case, based
on a Newtonian theory with a modified continuity equation (see
also \cite{Zanchin}).

We now discuss case (ii). Approximating equation (\ref{4.25}) for
$\omega x \gg 1$, $x \ll 1$, we get
\begin{equation} \label{4.29}
\hat{D}^{2}\Delta_R-\hat{D}\Delta_R+\frac{2}{3}\omega^{2}x^{2}\Delta_R+
\frac{\omega^{2}}{x^{2}}\Omega_{\Lambda}\Delta_R=0 \, .
\end{equation}
Using equation (\ref{4.18}), we can write $\Omega_{\Lambda}$ as:
\begin{equation} \label{4.30}
\Omega_{\Lambda}=\frac{\Lambda}{3H^{2}}=\Omega_{\Lambda\,{\rm eq}}\frac{2x^{4}}
{x+1+\Omega_{\Lambda\,{\rm eq}}(2x^{4}-x-1)}
\end{equation}
where the index ${\rm eq}$ refers to the time of matter radiation
equivalence. This expression can, in this case, be approximated to
$\Omega_{\Lambda}\simeq \Omega_{\Lambda\,{\rm eq}}\, 2x^{4}$, where we
took $\Omega_{\Lambda\,{\rm eq}}\ll 1$. As $\Omega_{\Lambda\,{\rm eq}}$ is
constant, equation (\ref{4.29}) can be solved to give
\begin{equation} \label{4.31}
\Delta_R=A\,\mathrm{exp}\,(\pm i \nu x) \, ; \quad
\nu^{2}=\Big(\frac{2}{3}+ 2\Omega_{\Lambda\,{\rm eq}}\Big)\omega^{2}\gg
1 \, .
\end{equation}
This represents a rapid oscillation at frequency $\nu$, which, due
to the fact that $\Omega_{\Lambda\,{\rm eq}}\ll 1$, is very close to the
frequency of oscillation of the solution for the $\Lambda=0$ case
(see \cite{padmanabhan1}). The solution for $\Delta_{DM}$, $\Delta_{DM}
=\textrm{ln}x$, is also
approximately that obtained for zero $\Lambda$.\\ \indent We
finally consider case (iii), $x\gg 1$. The equations are:
\begin{equation} \label{4.32}
\hat{D}^2\Delta_{DM}+\frac{1}{2}(1+3\Omega_{\Lambda})
\hat{D}\Delta_{DM}+\frac{3}{2}(\Omega_{\Lambda}-1)\Delta_{DM}=
\frac{16}{9}\omega^2 \Omega_{\Lambda} \Delta_R \, ,
\end{equation}
\begin{equation} \label{4.33}
\hat{D}^2 \Delta_R+\frac{1}{2}(-1+3\Omega_{\Lambda})\hat{D}
\Delta_R+\frac{2}{3}\omega^{2}x(1-\Omega_{\Lambda})\Delta_R=
\frac{3}{2}(1-\Omega_{\Lambda})\Delta_{DM}-\hat{D}\Delta_{DM} \, .
\end{equation}
For a value of $x$ sufficiently low, the perturbations will still
follow the behaviour they had in the $\Lambda=0$ case, as long as
the cosmological constant does not produce significant changes in
the dynamics of expansion. From (\cite{padmanabhan1}), we see that
the growing mode of $\Delta_{DM}$ is $Ax$, while the corresponding
mode of $\Delta_R$ is $(3A)/(4w^2)$. For $\Omega_\Lambda$ close to
1, the term in the first derivative in the left hand side of
(\ref{4.32}) is then close to $2Ax$, while the term in the right hand
side is $4A/3$; taking $x\gg1$, we can then ignore the term in
$\Delta_R$. Substituting $\Omega_\Lambda$ for $\Lambda/3H^2$, and
taking $H^2=H_0^2(\Omega_M\, a_0^3/a_{{\rm eq}}^3\,1/x^3
+\Omega_\Lambda)$, where the quantities $\Omega_M$ and
$\Omega_\Lambda$ now refer to their present values, we obtain for
the solutions to the equation in $\Delta_{DM}$:
\begin{equation} \label{4.34}
\delta_d\propto \frac{\sqrt{a_0^3\Omega_M+a_{{\rm eq}}^3\Omega_\Lambda
x^3}}{x^{3/2}} \, ,
\end{equation}
\begin{equation} \label{4.35}
\delta_g\propto x\,_2F_1\Big[\frac{1}{3},1,\frac{11}{6},-
\frac{\Omega_M}{\Omega_\Lambda}\frac{a_{{\rm eq}}^3}{a_0^3} x^3 \Big]\,.
\end{equation}
These correspond to the solutions obtained previously to the
equations for the perturbations in matter in the single fluid
relativistic case, (\ref{2.23}) and (\ref{2.24}).
We now take the limiting case $\Omega_\Lambda\simeq 1$. In this
limit, equations (\ref{4.32}) and (\ref{4.33}) become:
\begin{equation} \label{4.36}
\hat{D}^2\Delta_{DM}+2\hat{D}\Delta_{DM}=0 \, ,
\end{equation}
\begin{equation} \label{4.37}
\hat{D}^2 \Delta_R +\hat{D}\Delta_R+\hat{D}\Delta_{DM}=0 \, .
\end{equation}
The solution of (\ref{4.36}) will then be
\begin{equation} \label{4.38}
\Delta_{DM}=\delta_{DM}=A+Bx^{-2} \, ,
\end{equation}
and, taking the dominant mode in (\ref{4.38}), the solution of
(\ref{4.37}) is
\begin{equation} \label{4.39}
\Delta_R=\frac{3}{4}\delta_R=C+Dx^{-1} \, .
\end{equation}
As expected, the presence of a cosmological constant causes a halt
in the growth of perturbations, with the perturbations in dark
matter being greater than the radiation ones.

We can summarize the evolution of the perturbations as:
\begin{displaymath} \label{4.wl1}
\left\{ \begin{array}{lcll} \delta_{DM}=x^2 & ; & \delta_R=4/3 x^2
& \textrm{before entering the H.r.}
\\ \delta_{DM}=\textrm{ln} x & ; & \delta_R\, \textrm{oscillates} &
\textrm{radiation d.p., after entering the H.r.}
\\ \delta_{DM} & ; & \delta_R & \textrm{tend to a const. for
$\Lambda$ domination} \\ & & & \textrm{after
entering the H.r.}
\end{array} \right.
\end{displaymath}

\noindent\\ \emph{$\omega^2 \ll 1$: perturbations entering the
Hubble radius in the matter dominated phase}\\
We now treat the case $\omega^{2}\ll 1$. In this case, the modes
enter the Hubble radius in the matter dominated phase.  The
solutions for $x\ll 1$ are approximately the same as in the case
$\Lambda=0$, (see \cite{padmanabhan1}). We find the
solutions for the case $w^{2}x^{2}\gg 1$ and $x\gg 1$. As there is
a coupling (through terms of order $\omega x$) between $\delta$
and $S$ in equations (\ref{4.20}) and (\ref{4.21}), we can take
$\delta=S$ here, and, with these approximations, equations
(\ref{4.20}) and (\ref{4.21}) become:
\begin{equation} \label{4.26}
\hat{D}^{2}\delta+\bigg(\frac{1}{2}+
\frac{3}{2}\frac{\Lambda}{3H^{2}}\bigg)\hat{D}
+\frac{3}{2}\bigg(-1+\frac{\Lambda}{3H^{2}}\bigg)\delta=0 \, ,
\end{equation}
\begin{equation} \label{4.27}
\hat{D}^{2}S+\bigg(-\frac{1}{2}+
\frac{3}{2}\frac{\Lambda}{3H^{2}}\bigg)\hat{D}S=0
\, .
\end{equation}
In the $\Lambda$ dominated phase we can approximate
$\Omega_{\Lambda}\equiv\Lambda/(3H^{2})$ as a constant,
since $H$ will be approximately constant. The
dominant growing mode is then given by:
\begin{equation} \label{4.28}
\delta_{g}=S_{g}=x^{n_{g}} \, , \quad
n_{g}=\frac{1}{4}(-1-3\Omega_{\Lambda}+
\sqrt{25-18\Omega_{\Lambda}+9\Omega_{\Lambda}^2})
\, ;
\end{equation}
and the decaying mode by
\begin{equation} \label{4.28a}
\delta_{d}=S_{d}=x^{n_{d}} \, , \quad
n_{d}=\frac{1}{4}(-1-3\Omega_{\Lambda}-
\sqrt{25-18\Omega_{\Lambda}+9\Omega_{\Lambda}^2})
\, .
\end{equation}
For $\Omega_{\Lambda}=1$, these exponents become $n_g=0$ and
$n_d=-2$, which correspond to the solution (\ref{3.25}). For
$\Omega_{\Lambda}=0.7$, we have $n_g=0.25$ and $n_d=-1.8$. In the
limit $\Lambda\to 0$, these reduce to the appropriate solutions,
$n_{g}=1$ and $n_{d}=-3/2$. The dominant modes can then be summarized as
follows:
\begin{displaymath} \label{4.wg1}
\left\{ \begin{array}{lcll} \delta=x^2 & ; &S=\textrm{ln}x &
\textrm{before entering the H.r.} \\ \delta=x & ; & S=x &
\textrm{matter d.p., after entering the H.r.}
\\ \delta=x^{n_g} & ; & S=x^{n_g} & \textrm{limiting case for 
$\Lambda$ d.p.
after entering the H.r.}
\end{array} \right.
\end{displaymath}

\section{Newtonian Approximation} \label{secNewt}

\subsection{Equations}

\indent\indent The equations in the Newtonian approximation are
derived classically, so the introduction of a cosmological
constant will not change them; it will only influence the
solutions when we consider the dynamics of expansion. We will here
give a very brief description of the procedure and present the
equations we will be using. We start with the classic fluid
equations of mass conservation and pressure support:
\begin{equation} \label{2.1}
\dot{\rho} \equiv \frac{\partial \rho}{\partial t} + (\mathbf{v}
\cdot \nabla) \rho = -\rho (\nabla \cdot \mathbf{v})
\end{equation}
\begin{equation} \label{2.2}
\dot{\mathbf{v}} = -\nabla \phi - {\rho}^{-1} \nabla p
\end{equation}
We relate the velocity $v$ to the Hubble constant $H$ by Hubble's
Law, ${\mathbf{v}}(t,{\mathbf{x}})=H(t){\mathbf{x}}$, and consider
quantities to be of the form $A=A_b+\delta A$. We substitute these
in equations (\ref{2.1}) and (\ref{2.2}), and linearize them,
taking only the terms up to first order. Using $\nabla^{2} \delta
\phi=4 \pi G \delta \rho$ and the definitions $\delta p=c_{\rm s}^{2}
\delta \rho$ and $\delta=\delta\rho/\rho$, we arrive at the
following equation for the perturbations:
\begin{equation} \label{2.11}
\ddot{\delta} + 2 H_{b} \dot{\delta} - c_{\rm s}^{2} a^{-2} \nabla^{2}
\delta = 4 \pi G \rho_{b} \delta \, ,
\end{equation}
where the $a^{-2}$ term in the Laplacian comes from considering
Friedmann comoving coordinates. Considering perturbations with
wavelength $\lambda$, and taking the Fourier transform, we can
write the $\nabla^{2}$ term as $-k^{2}$. Equation (\ref{2.11})
then takes the form:
\begin{equation} \label{2.12}
\ddot{\delta} + 2 H_{b} \dot{\delta} +\frac{k^{2}c_{\rm s}^{2} }{a^{2}}
\delta = 4 \pi G \rho_{b} \delta \, .
\end{equation}
\indent This discussion can be generalized for the multi-fluid
case. The corresponding equation is then
\begin{equation} \label{2.13}
\ddot{\delta}_{N} + 2 H_{b} \dot{\delta}_{N} - {c_{\rm s}}_N^{2} a^{-2}
\nabla^{2} \delta_{N} = 4 \pi G \sum_{M} \rho_{M} \delta_{M}
\end{equation}
where the sum on the right hand side is over all components. A
smoothly distributed component (e.g. the cosmological constant
term) does not contribute to this term.

\subsection{Solutions}

\indent\indent We consider here the case where $\lambda \gg
\lambda_{J}$. We can then ignore pressure support, that is, ignore
the $k^{2}$ term in the left hand side of (\ref{2.12}), and get
the equation
\begin{equation} \label{2.14}
\ddot{\delta}+2H\dot{\delta}-4 \pi G \rho \delta=0
\end{equation}
As shown by Heath (\cite{Heath}), an integral expression for the growing
solution of equation (\ref{2.14}) for any value of $\Omega_M$ and
$\Omega_{\Lambda}$ can be obtained by noticing that $H(t)$ is
itself a decaying solution to this equation. The growing mode can
then be given by
\begin{equation} \label{2.20}
\delta\propto H(z) \int_{z}^{\infty}\frac{1+x}{H^3(x)} dx \, ,
\end{equation}
with
\begin{equation} \label{2.20a}
H(z)=H_{0}(1+z)\bigg[1+\Omega_{M}z+\Omega_{\Lambda}\Big(\frac{1}
{(1+z)^{2}}-1\Big)\bigg]^{1/2}
\end{equation}
and $z=a_0/a-1$ is the redshift.\\ \indent An analytical solution
for the $\Lambda=0$ case is widely known (see, for example,
\cite{padmanabhan1}), and an analytical expression for
(\ref{2.20}) in terms of elliptic integrals is given by Eisenstein
(\cite{Eisen}). A good approximation to this integral, given by
\begin{equation} \label{2.21}
\delta\approx a\, \frac{5}{2}\Omega_M
\bigg[\Omega_M^{4/7}-\Omega_{\Lambda}
+\bigg(1+\frac{1}{2}\Omega_M\bigg)\bigg(1+\frac{1}{70}\Omega_{\Lambda}
\bigg)\bigg]^{-1} \, ,
\end{equation}
(which is normalized to give $\delta=a$ for $\Omega_M=1\, ,
\,\,\Omega_{\Lambda}=0$) was given by Carroll et al.
(\cite{carrolletal}). Taking the limit $\Omega_M\simeq0$ with
$\Omega_{\Lambda}\neq1$, the integral (\ref{2.20}) can be
approximated by
\begin{equation} \label{2.21a}
\delta\propto \frac{1}{1-\Omega_{\Lambda}} \, ,
\end{equation}
which is a constant. We expect the perturbations to stop growing
when the cosmological constant becomes the dominant energy form,
because the expansion becomes faster than the gravitational
collapse of the perturbations. The expansion timescale is given
roughly by $t_{\rm exp}\sim(G\rho_{\rm
dominant})^{-1/2}\sim(G\rho_{\Lambda})^{-1/2}$, and is smaller
than the collapse timescale, $t_{\rm grav}\sim(G\rho_M)^{-1/2}$,
as long as $\rho_{\Lambda}>\rho_M$; thus, the rapid expansion of
the background prevents the growth of the perturbations by
gravitational collapse.
\\ \indent
We now study the perturbations in the matter component in a
universe with a cosmological constant, in the case
$\lambda\gg\lambda_{J}$ where we can ignore the pressure support.
The perturbation equation is simply (\ref{2.14}), with H given by
(\ref{friedmann}). In order to use the Newtonian equations, we are
considering here the case when the universe is well into the
matter dominated phase, and as such we can ignore the radiation
term in (\ref{friedmann}). Further, as we are not taking into account
the radiation, we need not distinguish between dark matter and
baryonic matter, as they do not have different interactions with
the cosmological constant term, unlike what happened with
radiation.

Using the variable $\overline{x}=a/a_{\Lambda}$, we get the
equation:
\begin{equation} \label{2.22}
(1+\overline{x}^{3})\overline{x}^{2}\delta\,''+\frac{3}{2}
(\overline{x}+2\overline{x}^{4})\delta\,
'-\frac{3}{2}\,\delta=0
\end{equation}
The solutions to these equations are the same as the ones obtained
earlier for the relativistic case, given by (\ref{2.23}) and
(\ref{2.24}).

\section{Conclusion}
\indent We have studied the equations for the gravitational
collapse of density perturbations in an expanding universe of the
Friedmann-Robertson-Walker type, whose dynamics is described by
equation (\ref{friedmann}), taking into account the presence of the
cosmological constant $\Lambda$ (or equivalently, considering the universe 
permeated by a quintessence fluid with an effective equation of state 
given by $p=-\rho$). We have confirmed that the
cosmological constant does not have much influence in the
behaviour of perturbations until it becomes a significant factor
in the expansion of the universe, see equation (\ref{friedmann}).
Thus, as we have explicitly shown for the particular case of
perturbations in radiation for the two-fluid case in (\ref{4.31}),
the growth of the perturbations for $\Lambda\neq0$ is not
significantly altered in the radiation dominated phase, when
compared to a $\Lambda=0$ universe. On the other hand, when the
cosmological constant becomes important in the expansion, at later
times, we have seen that significant differences arise in the
behaviour of perturbations. This is caused by the fact that the
expansion becomes faster than the gravitational collapse of the
perturbations, thereby preventing any further growth. This
behaviour can be seen, for example, in the exponent $n_g$ of the solution
(\ref{4.28}), which for $\Lambda=0$ is $n_g=1$, while for $\Omega_\Lambda=0.7$
we have $n_g=1/4$. A more detailed
solution for the transition from the phase of matter domination to
that of $\Lambda$-domination has been obtained, both in the fully
relativistic case and in the Newtonian approximation (which we
have shown to be a good approximation in this case). This solution
is given by equation (\ref{2.24}), and is illustrated in
Fig. 1. In this figure we can clearly see the
transition from the behaviour in the matter domination phase
$(x\ll1)$, when the perturbations grow as $a$, to the phase of
cosmological constant domination $(x\gg1)$, when the perturbations
stop growing altogether.

\bigskip
{\bf Acknowledgements:}
This work was partially funded by Funda\c c\~ao para 
a Ci\^encia e Tecnologia FCT through project ESO/PRO/1250/98.

\end{document}